\documentclass[prd,aps, tightenlines, preprintnumbers, showpacs, nofootinbib,superscriptaddress,notitlepage,11pt]{revtex4-1}

\usepackage{epsfig}
\usepackage{amsmath,amsthm,amssymb}
\usepackage{mathtools}
\usepackage{siunitx}

\usepackage[usenames,dvipsnames]{color}

\newcommand{\beq}{\begin{eqnarray}}
\newcommand{\eeq}{\end{eqnarray}}

\newcommand{\nn}{\nonumber \\}

\newif\iffigure
\figuretrue

\begin{document}

\title{Classical and quantum entropy of parton distributions}

\author{Yoshikazu Hagiwara}
\affiliation{Department of Physics, Kyoto University, Kyoto 606-8502, Japan}

\author{Yoshitaka Hatta}
\affiliation{Yukawa Institute for Theoretical Physics, Kyoto University, Kyoto 606-8502, Japan}

\author{Bo-Wen Xiao}
\affiliation{Key Laboratory of Quark and Lepton Physics (MOE) and Institute
of Particle Physics, Central China Normal University, Wuhan 430079, China}

\author{Feng Yuan}
\affiliation{Nuclear Science Division, Lawrence Berkeley National
Laboratory, Berkeley, CA 94720, USA}

\begin{abstract}
We introduce the semiclassical Wehrl entropy for the nucleon as a measure of complexity of the multiparton configuration in phase space. This gives a new perspective on the nucleon tomography. We evaluate the entropy in the small-$x$ region and compare with the quantum von Neumann entropy. We also discuss the growth of entropy at small-$x$ and argue that it eventually saturates due to the Pomeron loop effect. 
\end{abstract}
\pacs{24.85.+p, 12.38.Bx, 14.20.Dh}
\maketitle

\section{Introduction}

In the study of the  partonic structure of hadrons, the methods of thermodynamics and statistical physics have often turned out to be useful. 
For example, there exist parameterizations of the parton distribution functions (PDFs) inspired by the Bose-Einstein and Fermi-Dirac distributions  \cite{Mac:1989ki,Bhalerao:1996fc,Bourrely:2001du}. Another example is that the energy evolution of  the density of soft gluons has been treated as a reaction-diffusion process in statistical physics \cite{Munier:2003vc,Iancu:2004es}. 
 Intuitively, if there is a multitude of partons involved in a given process, it is reasonable to expect that certain features of observables admit a simple statistical description. The picture becomes increasingly more attractive at very high energy or in the small-$x$ region where the number of gluons grows rapidly.  

In recent years, several authors have introduced the notion of entropy of small-$x$ gluons in the hadron wavefunction and discussed its connection to the multiplicity in the final state \cite{Kutak:2011rb,Peschanski:2012cw,Kovner:2015hga,Kharzeev:2017qzs,Shuryak:2017phz,Berges:2017zws} (see also \cite{Miller:2003ci}). A hadron in its ground state is a pure quantum state for which the standard quantum (von Neumann) entropy vanishes. Yet, one can think of various types of partons with different values of $x$ as different subsystems which are entangled to each other. As experiment can only probe a small part of the hadron wavefunction above a certain value of $x$ while the rest being integrated, an entanglement entropy may be defined and calculated. 
However, previous discussions along this line relied on particular formalisms at small-$x$ (the Color Glass Condensate formalism \cite{Kovner:2015hga} and the dipole formalism \cite{Kharzeev:2017qzs}) which cannot be straightforwardly generalized to the quark sector or to the large-$x$ region. It would be interesting to have a more accessible, model-independent definition of entropy in terms of the quark and gluon field operators so that it can be analyzed by various perturbative and nonperturbative means.

In this paper, we study the entropy of quarks and gluons defined through the QCD Wigner \cite{wigner} and Husimi \cite{husimi} distributions. These are multi-dimensional phase space distributions of quarks and gluons inside a hadron, and have been actively pursued in the context of nucleon tomography \cite{Belitsky:2003nz,Meissner:2009ww,Lorce:2011kd,Hatta:2011ku,Lorce:2011ni,Mukherjee:2014nya,Hagiwara:2014iya,Hatta:2016dxp,Hagiwara:2016kam,Gutsche:2016gcd,Zhou:2016rnt,Bhattacharya:2017bvs,Hagiwara:2017fye,More:2017zqp,Raja:2017xlo}. In fact, it is quite natural to define an entropy via phase space distributions.    In quantum mechnics, the corresponding construction is known as the Wehrl entropy \cite{wehrl} which is a semiclassical counterpart of the fully quantum von Neumann entropy. The Wehrl entropy has been previously discussed in QCD in \cite{Kunihiro:2008gv,Tsukiji:2016krj,Tsukiji:2017pjx} for  a different purpose (the problem of thermalization in heavy-ion collisions) with a totally different definition. The present definition, appropriate for the study of the nucleon structure, was briefly given in \cite{Hatta:2015ggc}, but was not explored. Here we give a general discussion of the Wehrl entropy associated with the QCD Husimi distribution \cite{Hagiwara:2014iya} and evaluate it in the  small-$x$ region in a  model that features the gluon saturation effects. We also extend the result of \cite{Kharzeev:2017qzs} by including the so-called Pomeron loop effect and demonstrate that this leads to the saturation of entropy at small-$x$.

\section{Entropy in classical and quantum systems}

In this section we briefly review the definitions of entropy in classical and quantum mechanics. For simplicity, we consider a one-dimensional system, but generalization to arbitrary dimensions is straightforward. In statistical physics and kinetic theory,  entropy is defined via the phase space distribution function $f(q,p)$ ($q,p$ are the coordinate and momentum of particles) as    
\beq
S_{cl}=-\int \frac{ dq dp}{2\pi \hbar} f(q,p) \ln f(q,p). \label{cl}
\eeq
 If the system is out of equilibrium, $f$ depends on time $t$ according to the Boltzmann equation $\frac{\partial}{\partial t}f=C[f]$. As is well known, $S_{cl}(t)$ increases monotonically and eventually saturates as the system reaches equilibrium.

For a quantum system, the usual definition of entropy is the von Neumann entropy
\beq
S_{vN}=-{\rm Tr} \rho \ln \rho, \label{coh}
\eeq
where $\rho$ is the density matrix. For a pure state $\rho=|\psi\rangle \langle \psi|$, $S_{vN}$ vanishes. A nonzero $S_{vN}$ then  measures the degree of deviation from  a pure state.  For a density matrix  of the form $\rho=\sum_n p_n|\psi_n\rangle \langle \psi_n|$, it is given by
\beq
S_{vN}=-\sum_n p_n \ln p_n \label{p}.
\eeq

The two entropies $S_{cl}$ and $S_{vN}$ are not simply related to each other. In particular, the latter does not reduce to the former in the limit  $\hbar \to 0$. To make a connection between the two, Wehrl introduced an intermediate definition of entropy \cite{wehrl}. Let $|\lambda\rangle$ with $\lambda=\frac{1}{\sqrt{2\hbar}}(q+ip)$ be the coherent state which is the eigenstate of the annihilation operator $a|\lambda\rangle =\lambda |\lambda\rangle$. Taking the trace in (\ref{coh}) in the coherent state basis, one gets
\beq
S_{vN}=-\int \frac{dqdp}{2\pi \hbar} \langle \lambda|\rho \ln \rho |\lambda\rangle.
\eeq
The Wehrl entropy is obtained by replacing  $\langle \lambda|\rho \ln \rho |\lambda\rangle$ with $\langle \lambda|\rho|\lambda\rangle  \ln \langle \lambda|\rho |\lambda\rangle$. Introducing the Husimi distribution \cite{husimi}, 
\beq
H(q,p)=\langle \lambda |\rho |\lambda\rangle,
\eeq
one can write the Wehrl entropy as 
\beq
S_{W}=-\int\frac{dqdp}{2\pi \hbar} H(q,p)\ln H(q,p). \label{sh}
\eeq
Since the function $f(x)=-x\ln x$ is concave, it follows that 
\beq
S_{W} > S_{vN} \ge 0.   \label{in}
\eeq
The equality $S_{W}=S_{vN}$ is impossible \cite{wehrl}, and this means that $S_W$ is always nonzero even for a pure state. 

The Husimi distribution $H(q,p)$ is the closest analog in quantum mechanics of the classical phase space distribution $f(q,p)$. One may be tempted to use instead the more well-known  Wigner distribution \cite{wigner}
\beq
W(q,p)= \int_{-\infty}^\infty dy e^{-ipy/\hbar}\langle q+y/2|\rho|q-y/2\rangle,
\eeq
and define 
\beq
\widetilde{S}_W  \equiv -\int\frac{dqdp}{2\pi \hbar} W(q,p)\ln W(q,p). \label{sw}
\eeq
However, $W(q,p)$ is not positive definite, and therefore its logarithm is not well defined everywhere. 
 These two distributions are related by Gaussian smearing 
\beq
H(q,p) = \frac{1}{\pi \hbar}\int dq'dp' e^{-(q-q')^2/\hbar -(p-p')^2/\hbar} W(q',p').
\eeq
As is clear from this expression, the Husimi distribution smooths out localized fluctuations in a phase space volume $\Delta q \Delta p < \hbar/2$. Such fluctuations are unphysical in that they do not bring about measurable consequences due to the uncertainty principle. Due in part to this loss of information, $S_W$ is always nonvanishing.

As an example, consider a one-dimensional harmonic oscillator with the Hamiltonian $H=\frac{p^2+q^2}{2}$. For the $n$-th excited state, the  Husimi distribution can be analytically computed as
\beq
 H(q,p) = \frac{1}{n!} e^{-H/\hbar} \left(\frac{H}{\hbar}\right)^n.
\eeq
Substituting this into (\ref{sh}), we find 
\beq
S_W = n+1 +\ln n! - n\psi(n+1), \label{har}
\eeq
where $\psi$ is the digamma function. Asymptotically, $S_W\approx \frac{1}{2}\ln n$. On the other hand, the Wigner distribution oscillates and becomes negative  (except for the ground state $n=0$). Thus the alternative definition (\ref{sw}) does not make sense. 
Note that the von Neumann entropy vanishes for all levels $n$ because they are pure states.

\section{Wehrl entropy in QCD}

Let us now consider the entropy of partons in 1+3 dimensional QCD. Since quantum field theory is not commonly formulated in terms of state vectors $|\psi\rangle$ and a density matrix,  it appears difficult to define and evaluate the von Neumann (entanglement) entropy $S_{vN}$ in a model-independent way. (Such a construction is nevertheless possible within certain frameworks \cite{Miller:2003ci,Kovner:2015hga,Kharzeev:2017qzs}, and we shall discuss one such model in a later section.) We thus turn to the Wehrl entropy. The QCD Wigner distribution $W(x,b_\perp,k_\perp)$ is the generalization of the collinear PDF to include dependences on transverse momentum $k_\perp$ and impact parameter $b_\perp$. 
They are defined by ($\hbar=1$ in the following) 
\beq
xW^{\pm}_q(x,b_\perp,k_\perp) &=& \int \frac{dz^- d^2z_\perp}{(2\pi)^32P^+}\int \frac{d^2\Delta_\perp}{(2\pi)^2} e^{-ixP^+z^- -iq_\perp\cdot z_\perp} \nn
&& \times \left\langle P+\Delta_\perp/2 \right|\bar{q}(b_\perp+z/2)U^{\pm} q(b_\perp -z/2)\left|P-\Delta_\perp/2\right\rangle,
\eeq
for quarks and 
\beq
xW^{\pm\pm}_g(x,b_\perp,k_\perp) &=& \int \frac{dz^- d^2z_\perp}{(2\pi)^3P^+}\int \frac{d^2\Delta_\perp}{(2\pi)^2} e^{-ixP^+z^- -iq_\perp\cdot z_\perp} \nn
&& \times \left\langle P+\Delta_\perp/2 \right|{\rm Tr}[ F^{+\alpha}(b_\perp+z/2)U^{\pm} F^{+}_{\ \alpha}(b_\perp -z/2) U^{\pm} ]\left|P-\Delta_\perp/2\right\rangle,
\eeq
for gluons. $|P\rangle$ is the single hadron state (usually the proton) with momentum $P^\mu$.  $U^\pm$ is the staple-shaped fundamental Wilson line along the light-cone extending to $z^-=\pm \infty$. In the quark case, $W^+_q$ and $W^-_q$ are simply related by $PT$ transformation. In the gluon case, there are two distinct Wigner distributions, the Weisz$\ddot{{\rm a}}$cker-Williams (WW) distribution $W_{WW}=W_g^{++}$ and the dipole Wigner distribution $W_{dip}=W_g^{+-}$. The difference in the Wilson line configuration means that they contribute to different observables. For instance, $W_{dip}$ contributes to diffractive dijet production in $ep$ and $pA$ collisions \cite{Hatta:2016dxp,Hagiwara:2017fye}.

The QCD Husimi distributions for quarks and gluons are defined by smearing the corresponding Wigner distributions in phase space $(b_\perp,k_\perp)$   \cite{Hagiwara:2014iya}
\beq
xH_q(x,b_\perp,k_\perp) = \frac{1}{\pi^2} \int d^2b'_\perp d^2k'_\perp e^{-(b_\perp-b'_\perp)^2/\ell^2 -\ell^2 (k_\perp-k'_\perp)^2}xW_q(x,b'_\perp,k'_\perp),
\eeq
\beq
xH_{WW/dip}(x,b_\perp,k_\perp) = \frac{1}{\pi^2} \int d^2b'_\perp d^2k'_\perp e^{-(b_\perp-b'_\perp)^2/\ell^2 -\ell^2 (k_\perp-k'_\perp)^2}xW_{WW/dip}(x,b'_\perp,k'_\perp), \label{f}
\eeq
 where $\ell$ is an arbitrary parameter with the dimension of length. Notice that the widths of the two Gaussians are inversely related such that the smearing is done in the minimum uncertainty region $\Delta b_\perp \Delta k_\perp = \frac{1}{2}$.
 
 Unlike in quantum mechanics, a general proof of positivity of the QCD Husimi distributions $H_{q/WW/dip}$ is unfortunately not available.  However, Refs.~\cite{Hatta:2015ggc,Hagiwara:2016kam} provided nontrivial examples in which the Husimi distribution is smooth and positive  everywhere although the corresponding Wigner distribution oscillates between positive and negative values. (We shall see another example of this below.) We thus assume the positivity of the Husimi distribution as a working hypothesis and define the Wehrl entropy as a function of $x$ 
  \beq
S_W(x)&\equiv &-\int d^2b_\perp d^2 k_\perp xH(x,b_\perp,k_\perp)\ln xH(x,b_\perp,k_\perp),  \label{def}
\eeq
 where $H=H_{q/WW/dip}$. We opt to use $xH$ instead of $H$ since we are considering entropy  per unit rapidity $Y=\ln 1/x$.
If the Wigner distribution turns out to be positive definite in some approximations or model calculations, we may as well define an entropy by
  \beq
\widetilde{S}_W(x)&\equiv &-\int d^2b_\perp d^2 k_\perp xW(x,b_\perp,k_\perp)\ln xW(x,b_\perp,k_\perp). \label{def2}
\eeq
 However, such a definition has limited applicability. Firstly, the Wigner distribution typically  has a perturbative tail $W\sim 1/k_\perp^2$ which makes the $k_\perp$ integral logarithmically divergent. While one may argue that this should be cut off by the resolution scale $Q^2$,  a more serious problem is that most likely positivity is not preserved by the QCD evolution \cite{Mukherjee:2014nya,Hagiwara:2014iya}. 
 
In the following, we arbitrarily neglect the overall prefactor of $xH$ and $xW$ inside the logarithm. This factor modifies the entropy only by an amount proportional to the collinear parton distribution function (PDF) $\int d^2b_\perp d^2k_\perp xH_{q,g}(x,b_\perp,k_\perp) =\int d^2b_\perp d^2k_\perp xW_{q,g}(x,b_\perp,k_\perp)  = xq(x),xg(x)$. It thus does not carry nontrivial information about the phase space structure of partons.

As a trivial example, consider a free electron or a quark moving in the positive $z$-direction.  The Wigner and Husimi distributions are (setting $x=1$)
\beq
W(b_\perp,k_\perp)= \delta^{(2)}(b_\perp)\delta^{(2)}(k_\perp), \qquad 
H(b_\perp,k_\perp)=\frac{e^{-b_\perp^2/\ell^2 - \ell^2 k_\perp^2}}{\pi^2}.
\eeq
 While the Wigner distribution is positive, its logarithm does not make sense. The Wehrl entropy can be evaluated from the Husimi distribution as 
\beq
S_W= \frac{1}{\pi^2} \int d^2b_\perp d^2k_\perp e^{-b_\perp^2/\ell^2 - \ell^2 k_\perp^2} \left(\frac{b_\perp^2}{\ell^2} + \ell^2 k_\perp^2\right) = 2.
\eeq
The nonvanishing value reflects our inability to precisely determine  position and momentum simultaneously due to the uncertainty principle.

\section{Wehrl entropy of small-$x$ gluons}

In this section, we shall focus on the Wehrl entropy generated by gluons in the small-$x$ region, or equivalently, large rapidity region $Y=\ln \frac{1}{x} \gg 1$.  As already mentioned in the introduction, the number of gluons grows rapidly as $x$ is decreased, and these gluons show collective behaviors which may be treated semiclassically. It is thus very interesting to consider the Wehrl entropy of such states. 
 
As recently shown in \cite{Hatta:2016dxp}, at small-$x$ the dipole Wigner distribution takes the following simple form  
\beq
xW_{dip}(x,b_\perp, k_\perp)= \frac{2N_c}{\alpha_s (2\pi)^2} \int \frac{d^2r_\perp}{(2\pi)^2}e^{ik_\perp \cdot r_\perp}\left(\frac{1}{4}\nabla_{b_\perp}^2+k_\perp^2\right) {\cal S}(x,b_\perp,r_\perp), \label{d1}
\eeq
where ${\cal S}$ is the forward S-matrix of a dipole of size $r_\perp$ at impact parameter $b_\perp$ scattering off the hadron of interest.  In general, $W_{dip}$ is not positive definite due to the $b_\perp$-derivative term.  
An equally simple, general expression of the WW Wigner distribution is not available, but in a quasiclassical approximation one can deduce the following form \cite{Kovchegov:1998bi, Dominguez:2011wm}
\beq
xW_{WW}(x, b_\perp,k_\perp) = \frac{N_c^2-1}{4\pi^4 \alpha_s N_c}\int d^2r_\perp e^{ik_\perp \cdot r_\perp} \frac{1}{r_\perp^2}  \left(1-\tilde{\cal S}\right), \label{www}
\eeq
 where again $\tilde{\cal S}=\tilde{\cal S}(x,b_\perp,k_\perp)$ is the dipole S-matrix in the adjoint representation. After integrating over $b_\perp$, one can reproduce the  WW unintegrated gluon distribution \cite{Kovchegov:1998bi, Dominguez:2011wm}.

Let us evaluate these expressions in a GBW-like model \cite{GolecBiernat:1998js}
\beq
{\cal S}=e^{-\frac{1}{4}r^2_\perp Q^2_s(x,b_\perp)} , \quad \tilde{\cal S}=e^{-\frac{1}{4}r^2_\perp Q^2_{sg}(x,b_\perp)}, \label{gbw}
\eeq
where $Q_{s(g)}$ is the quark (gluon) saturation momentum which we assume to be of the form 
 $Q_{s(g)}^2(x,b_\perp) = \Lambda^2\left(\frac{1}{x}\right)^\alpha h_{(g)}(b_\perp^2 \Lambda^2)$. ($\Lambda$ is the confinement scale.) 
 Inserting (\ref{gbw}) into (\ref{www}), we find 
\beq
xW_{WW}(x, b_\perp,k_\perp) &=& \frac{N_c^2-1}{4\pi^4\alpha_s N_c}\int \frac{d^2r_\perp}{r_\perp^2} e^{ik_\perp \cdot r_\perp} \left(1-e^{-\frac{1}{4}Q_{sg}^2(x,b_\perp)r_\perp^2}\right) \nn
&=& \frac{N_c^2-1}{4\pi^3 \alpha_s N_c}\Gamma\left[0,\frac{k_\perp^2}{Q_{sg}^2(x,b_\perp)}\right], \label{not}
\eeq
where $\Gamma[0, z]$ is the incomplete Gamma function. Note that (\ref{not}) is positive definite. While this may seem natural in view of the fact that the  WW unintegrated gluon distribution  admits a probabilistic interpretation, we emphasize that the positivity of $W_{WW}$ is not guaranteed in general and likely to be violated by the quantum evolution.  
 Anyway, since (\ref{not}) is positive, we can adopt the simpler definition (\ref{def2}) and obtain
\beq
\widetilde{S}_W&\equiv &-\int d^2b_\perp d^2 k_\perp xW(x,b_\perp,k_\perp)\ln xW(x,b_\perp,k_\perp)  \nn
&\simeq&- \frac{N_c^2-1}{4\pi^3 \alpha_s N_c}\int d^2k_\perp d^2b_\perp \Gamma\left[0,\frac{k_\perp^2}{Q_{sg}^2(x,b_\perp)}\right] \ln  \Gamma\left[0,\frac{k_\perp^2}{Q_{sg}^2(x,b_\perp)}\right] \nn
&=& 0.248\frac{N_c^2-1}{4\pi \alpha_s N_c}\int_0^{\infty} db^2_\perp Q_{sg}^2(Y,b^2_\perp). 
\eeq
We see that the entropy grows exponentially in rapidity $\widetilde{S}_W\sim Q_s^2(Y)\sim e^{\alpha Y}$ in this model. This is essentially due to the transverse dynamics of QCD, and is also a consequence of geometric scaling which holds perfectly for the model at hand (that is, $xW(x,q_\perp)$ depends only on the ratio $q^2_\perp/Q_s^2(Y)$). The parametric dependence $\widetilde{S}_W\propto C_FQ_s^2/\alpha_s$ agrees with the previous results in \cite{Kutak:2011rb,Peschanski:2012cw,Kovner:2015hga} using other definitions of entropy. 
For a large nucleus with atomic number $A$, $\widetilde{S}_W\sim \int d^2b_\perp Q_s^2 \propto A^{2/3}A^{1/3}=A$. This is because the number of gluons in a large nucleus is additive in the quasiclassical approximation \cite{Mueller:1999yb}, and is consistent with the fact that the entropy is an extensive variable.

Next we turn to the dipole Wigner distribution. Using the same Gaussian ansatz (\ref{gbw}), it is evaluated as
\beq
xW_{dip}(x,b_\perp, k_\perp)= \frac{2N_c}{\alpha_s (2\pi)^2} \left( \frac{\partial}{\partial b_\perp^2} b_\perp^2 \frac{\partial}{\partial b_\perp^2}+k_\perp^2\right) \frac{e^{-\frac{k_\perp^2}{Q_s^2}}}{\pi Q_s^2}. \label{ab}
\eeq
For realistic profile functions $Q^2_s(b_\perp)$, we find that (\ref{ab}) is not positive definite.  This is related to the fact that the dipole distribution does not have a probabilistic interpretation due to the nontrivial gauge link dependence.   We thus compute instead the Husimi distribution (\ref{f}) 
\beq
xH(x,b_\perp,k_\perp) &=& \frac{2N_c}{\ell^4 \alpha_s  \pi^2 (2\pi)^2}\int d^2b'_\perp e^{-(b_\perp-b'_\perp)^2/\ell^2 -\frac{\ell^2}{1+\ell^2Q_s^2}k_\perp^2} \nn 
&& \times \left[\frac{(b_\perp -b'_\perp)^2}{\ell^2} + \frac{(\ell^2 Q_s^2)^2}{(1+\ell^2 Q_s^2)^2}\ell^2 k_\perp^2 -\frac{1}{1+\ell^2 Q_s^2}\right] \frac{\ell^2 }{1+\ell^2 Q_s^2}. \label{h}
\eeq
For an arbitrary function $Q^2_s(b_\perp)$ which is monotonically decreasing with increasing $b_\perp$, (\ref{h}) is positive definite. To see this, note that $\frac{1}{1+\ell^2 Q_0^2} \le \frac{1}{1+\ell^2 Q_s^2}<1$ where $Q_0^2 \equiv Q_s^2(b_\perp=0)$. We then find 
\beq
xH(x,b_\perp,k_\perp) > \frac{2N_c e^{-\ell^2 k_\perp^2}}{\ell^2 \alpha_s\pi^2 (2\pi)^2(1+\ell^2 Q_0^2)} \int d^2b'_\perp e^{-\frac{(b_\perp-b'_\perp)^2}{\ell^2}} \left[\frac{(b_\perp-b'_\perp)^2}{\ell^2}-1\right]=0.
\eeq
 We can thus safely compute the Wehrl entropy (\ref{def}). For large values of $Q_s^2$, $xH(x,b_\perp,k_\perp)$ depends on $k_\perp$ only through the ratio $k_\perp^2/Q_s^2$. It is then clear that the entropy behaves as $S_W \propto N_cQ_s^2/\alpha_s \sim e^{\alpha Y}$. 

Before leaving this section we note that one can also consider the entropy of small-$x$ {\it quarks}. The sea quark distribution has been computed in the small-$x$ formalism \cite{McLerran:1998nk,Mueller:1999wm,Marquet:2009ca}.  In a quasi-classical approximation, one can introduce the $b_\perp$-dependence in these results as
\beq
xW_q(x,b_\perp,k_\perp) = \frac{N_c}{4\pi^4} \int d^2k_{g\perp} F(x,k_{g\perp},Q_s^2(b_\perp))
\left(1-\frac{k_\perp \cdot (k_\perp-k_{g\perp})}{k_\perp^2-(k_\perp-k_{g\perp})^2}\ln \frac{k_\perp^2}{(k_\perp-k_{g\perp})^2}\right), 
\eeq
 where $F(k_{g\perp})$ is the Fourier transform of ${\cal S}(r_\perp)$. As already mentioned, this has a perturbatve tail $W_q\sim 1/k_{\perp}^2$. One can eliminate this tail by switching to the Husimi distribution and find $S_W \propto N_c Q_s^2$.  Thus the entropy of quarks is smaller than that of gluons by a factor of $\alpha_s$, as expected.

\section{Towards the  saturation of entropy}

The results in the previous section suggest that entropy grows indefinitely as the rapidity $Y=\ln 1/x$  is increased. However we do not believe that this rapid growth continues forever. In this respect it may be useful to draw an analogy to the classical entropy Eq.~(\ref{cl}) of a time-dependent system. $S_{cl}(t)$ grows monotonically and eventually reaches a plateau as the system equilibrates. When this occurs,  the collision term of the Boltzmann equation vanishes because  the `gain' terms are exactly canceled by the `loss' terms. In QCD, the rapidity $Y=\ln 1/x$ plays the role of time, and the rapid growth of entropy with $Y$ is essentially because one has included only the gain terms, namely, gluon splittings. By including the loss terms, or gluon recombinations, the number of gluons eventually saturates, and so does the entropy.

Unfortunately,   a complete treatment of both the splitting and recombination effects, or the {\it Pomeron loop} effect, is an unsolved open question. Here we adopt a simple 1+0 dimensional model which was originally introduced in the context of Mueller's dipole model \cite{Mueller:1994gb}.
Since there is no transverse phase space in 1+0 dimensions, the Wehrl entropy cannot be defined.  Still, in this model one can naturally introduce the density matrix and calculate the von Neumann entropy as was done recently in \cite{Kharzeev:2017qzs}. We thus study the effect of Pomeron loops on the von Neumann entropy with the hope of gaining some insights into the fate of entropy in actual QCD.

  Let $P_n(Y)$ be the probability to find $n$ dipoles (gluons) at time $Y$ starting from a single dipole at $Y=0$. $P_n$ satisfies the equation
\beq
\frac{d}{dY} P_n = -\alpha n P_n + \alpha(n-1) P_{n-1}. \label{pn}
\eeq 
This equation only describes gluon splittings, and $\alpha>0$ is the corresponding probability.  
Defining the generating function 
\beq
Z(Y,u)=\sum_{n=1}^\infty P_n(Y)u^n, \label{gen}
\eeq
one can show that (\ref{pn}) is equivalent to the following equation
\beq
\frac{d}{dY} Z = \alpha(Z^2-Z), \qquad Z(Y=0)=u,\label{uni}
\eeq
which can be easily solved as 
\beq
Z=\frac{u}{u+(1-u)e^{\alpha Y}}.
\eeq 
It immediately follow that
\beq
P_n = \left.\frac{1}{n!}\frac{d^n Z}{du^n} \right|_{u=0} = e^{-\alpha Y} (1-e^{-\alpha Y})^{n-1}.
\eeq
The von Neumann entropy for this system is defined (see (\ref{p})) and calculated as 
\beq
S_{vN} = -\sum_n P_n \ln P_n \approx \alpha Y = \ln \langle n\rangle,  \label{dipv}
\eeq 
at large-$Y$. 
Thus the entropy grows linearly with  $Y$, or equivalently, logarithmically with the average multiplicity in this model \cite{Kharzeev:2017qzs}. 

Generalization to 1+3 dimensional QCD is significantly more complicated. The $n$-dipole probability $P_n$ now depends on $n$ two-dimensional vectors $\{z_\perp\}$ specifying the size and impact parameter  of dipoles. 
 One then defines
\beq
S_{vN} = -\sum_n^\infty \int \prod_i^n d^2z_{\perp i} P_n(Y,\{z_\perp\}) \ln P_n(Y,\{z_\perp\}).
\label{von2}
\eeq
It seems very hard to evaluate this in full generality. In \cite{Kharzeev:2017qzs}, the authors used several approximations  and obtained a result  $S_{vN} \sim (\ln Q_s^2(Y)) Y \propto Y^2$, which however does not agree with the  behavior $S_W\sim Q_s^2 \sim e^Y$ found in  the previous section and also in \cite{Kutak:2011rb,Peschanski:2012cw,Kovner:2015hga}.  
 This is simply due to different definitions of entropy. Nevertheless, it is interesting to point out a structural similarity between the two definitions: In the dipole approach, and in a frame in which the target dipole is slowly moving, the S-matrix and $P_n$ are linearly related as (see, e.g., \cite{Hatta:2006hs})
\beq
{\cal S}(Y,b_\perp,r_\perp) = \sum_n \int \prod_i^n d^2z_{\perp i} P_n(Y,\{z_{\perp}\})s_0s_1\cdots s_n,
\eeq
where $s_i=s_i(z_{\perp i},z_{\perp i+1})$, with  $z_{\perp 0}=b_\perp+\frac{r_\perp}{2}$, $z_{\perp n}=b_\perp-\frac{r_\perp}{2}$,  is the S-matrix of an elementary dipole off the target.  Therefore, roughly we have $S_W \sim {\cal S}\ln {\cal S} \sim P_n\ln P_n$. The difference, then, appears to be attributed to an additional integration over the impact parameter  $b_\perp$ in (\ref{def}) which, by dimensional reasons, brings in a factor $Q_s^2$.\footnote{In Ref.~\cite{Kharzeev:2017qzs}, it was assumed that $P_n(\{z_\perp\})$ was a function only of the absolute value of dipole sizes $|z_{\perp i} - z_{\perp i+1}|$ and not of their impact parameter.}

Returning to the 1+0 dimensional problem, we now discuss the saturation of entropy by including the recombination effect.\footnote{To avoid confusion, we note that the saturation and unitarity of scattering amplitudes is achieved already in the model (\ref{pn}) (and its 1+3 dimensional generalization) \cite{Mueller:1994gb},  as suggested by the nonlinear term in (\ref{uni}). This is a consequence of the `duality' of high energy evolution: a splitting in the projectile can be viewed as a recombination in the target. The genuine recombination effect in the projectile is missing. } 
Following  \cite{Shoshi:2005pf,Bondarenko:2006rh}, we generalize (\ref{pn}) as 
\beq
\frac{d}{dY} P_n = -\alpha n P_n + \alpha(n-1) P_{n-1} + \beta n(n+1)P_{n+1} -\beta n(n-1)P_{n}, \label{new}
\eeq
with $\beta = \alpha_s^2\alpha >0$ and $\alpha_s^2\ll 1$. 
The last two terms represent the $2\to 1$ recombination process with probability $\beta$.  
As shown in \cite{Bondarenko:2006rh}, Eq.~(\ref{new}) admits a stationary solution ($P_n$ independent of $Y$) which is Poissonian. 
\beq
P_n = \frac{N^n}{n!}e^{-N}, \label{po}
\eeq
where $N\equiv 1/\alpha_s^2 =\langle n\rangle$. 
This already indicates that the entropy will saturate eventually. In order to study the preasymptotic behavior, it is convenient to consider the moments
\beq
 n^{(k)} \equiv \sum_n n(n-1)\cdots (n-k+1)P_n = \left. \frac{d^{k}}{d u^k} Z(u)\right|_{u=1},
\eeq
 where $Z$ is as defined in (\ref{gen}). 
The equation for $n^{(k)}$ reads
\beq
\frac{d}{dY} n^{(k)} = k\alpha n^{(k)} + k(k-1) \alpha n^{(k-1)} -k\beta n^{(k+1)} -\beta k(k-1)n^{(k)}\,.
\eeq
An approximate perturbative solution for $n^{(k=1)}$, neglecting terms of order  ${\cal O}(\alpha_s^2 \alpha Y)$ and ${\cal O}(e^{-\alpha Y})$, has been obtained in \cite{Shoshi:2005pf} (see Eq.~(13) there).
 It is straightforward to  generalize this result to arbitrary $k$. We find
\beq
  n^{(k)} \approx \frac{e^{k\alpha Y}}{\Gamma(k)}\sum_{i=0}^\infty (-X)^i \frac{\Gamma(k+i+1)\Gamma(i+k)}{\Gamma(i+1)}, \label{a}
\eeq
where $X\equiv \alpha_s^2 e^{\alpha Y}$. Physically, the index $i$ represents the number of Pomeron loop insertions.  The series (\ref{a}) has zero convergence radius, but is Borel summable. 
Using the identity $\Gamma(k)= \int_0^\infty dz z^{k-1}e^{-z}$, we can cast the above equation into 
\beq
n^{(k)} \approx e^{k\alpha Y}\int_0^\infty dz \frac{z^k}{(1+zX)^k}e^{-z}.
\eeq
This allows us to reconstruct the generating function
\beq
Z(u)= \sum_{k=0}^\infty \frac{1}{k!}n^{(k)}(u-1)^k = \int_0^\infty dz \, e^{-z+\frac{NzX}{1+zX}(u-1)},
\eeq
where $N\equiv \frac{1}{\alpha_s^2}$. We thus arrive at 
\beq
P_n = \left. \frac{1}{n!}\frac{d^n Z}{du^n}\right|_{u=0} &=& \frac{N^n}{n!}\int_0^\infty dz e^{-z-\frac{NzX}{1+zX}}\left[\frac{zX}{1+zX}\right]^n.
\label{sad} 
\eeq
\begin{figure}[h]
 \includegraphics[height=61mm]{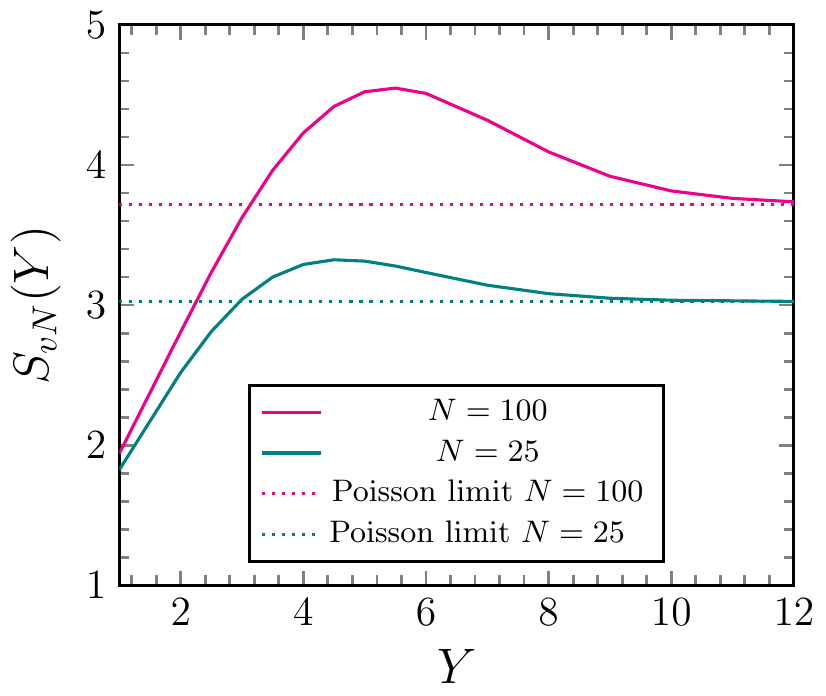} \includegraphics[height=60mm]{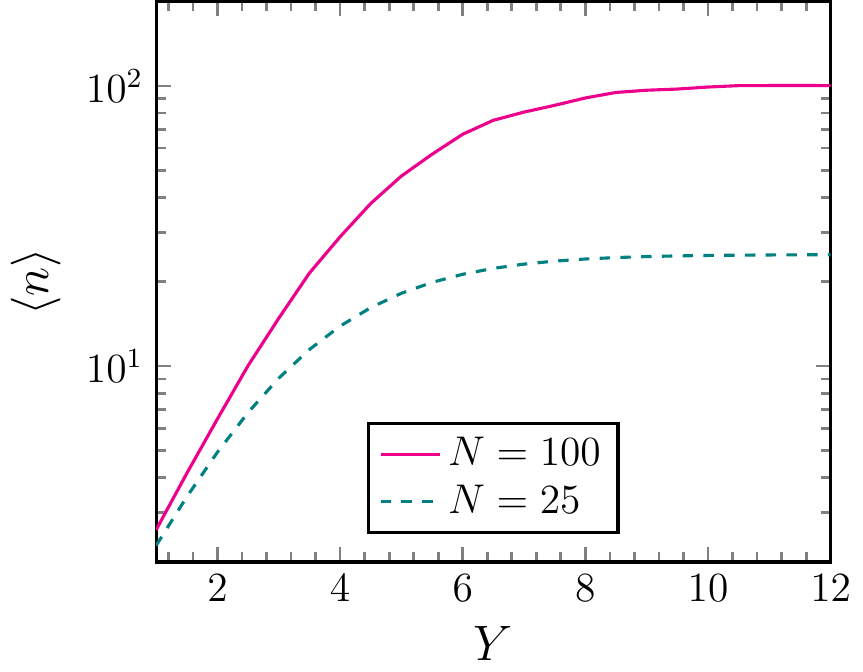}
\caption{von Neumann entropy $S_{vN}$ and $\langle n\rangle$ as a function of rapidity $Y$. We have set $\alpha=1$ and used (\ref{sad}) in (\ref{dipv}). The region $\alpha Y\lesssim 1$ is excluded because  our approximatios which led to (\ref{sad}) are not valid there.   
} 
\label{fig2}
\end{figure}
It is easy to check that $\sum_{n=0}^\infty P_n=1$. In Fig.~\ref{fig2} we show the von Neumann entropy numerically computed from (\ref{sad}).   
Interestingly, the entropy is not monotonic although the average multiplicity $\langle n\rangle$ is. This indicates that $S_{vN}$ depends not only on $\langle n \rangle$ but also  on higher moments.    $S_{vN}(Y)$ takes a maximal value when $X=\alpha_s^2 e^{\alpha Y}\sim 1$ and then starts to decrease and eventually saturates as $Y\to \infty$ to the known value of the Poisson distribution
\beq
S_{vN} \approx \frac{1}{2}\ln (2\pi eN) -\frac{1}{12N} + {\cal O}(1/N^2).  \label{limit}
\eeq
 Note that the coefficient in front of the logarithm $S_{vN}\sim c \ln \langle n\rangle$ has changed from $c=1$ in (\ref{dipv}) to $c=1/2$ (see also (\ref{har})), and this largely accounts for the decrease of  entropy  at $X>1$.\footnote{Incidentally, the aforementioned peak can disappear when $N<2\pi e$, since $ \frac{1}{2}\ln (2\pi eN)$ becomes larger than $\ln N$. We only focus on the case where $N$ is sufficiently large.}  

This non-monotonic behavior of entropy seems counterintuitive at first sight, but after all there is no `H-theorem' for this quantity.\footnote{A similar non-monotonic behavior of the Wehrl entropy has been observed in a different context \cite{Tsukiji:2016krj,Tsukiji:2017pjx}. }  Given that the von Neumann entropy measures the deviation from a pure state, we may say that the saturated gluon states probed at asymptotic energy are more ordered,
 in analogy to the terminology commonly used in condense matter physics. In other words, the transition to the saturation resembles a phase transition, although it may not be a genuine phase transition, since we are considering the quantum fluctuations of partons inside the wavefunction of a confined hadron. It is also interesting to note that the  Poisson distribution is known as the maximum entropy distribution among the $\infty$-generalized binomial distributions with fixed mean $\langle n\rangle$.

At last, we would like to comment on some observation related to the above probability distribution in Eq.~(\ref{sad}). When $X\to \infty$, $P_n$ reduces to the Poisson distribution (\ref{po}), as expected, since it is the stationary fixed point of Eq.~(\ref{new}). As long as $NX=e^{\alpha Y}$ is sufficiently large, we can also take the limit $X \ll 1$ and see that Eq.~(\ref{sad}) then reduces to 
\beq
P_n =\left(\frac{NX}{1+NX}\right)^n\frac{1}{1+NX},
\eeq
which is known as the geometric distribution with $\langle n\rangle =NX=e^{\alpha Y}$. In phenomenology, the so-called negative binomial distribution (NBD), which is defined with two parameters  $\langle n\rangle$ and $k$ as 
\beq
P_n^{NB} =\frac{\Gamma[n+k]}{\Gamma[n+1]\Gamma[k]}\left(\frac{\langle n \rangle}{k+\langle n \rangle}\right)^n\left(\frac{k}{k+\langle n \rangle }\right)^k,
\eeq
is often used to describe the multiplicity distribution in high energy collisions \cite{Alner:1985zc, GrosseOetringhaus:2009kz}, and it can be derived from the small-$x$ framework \cite{Gelis:2009wh, Dumitru:2012tw}. The geometric distribution is simply the special case of the NBD with $k=1$. It is also interesting to notice that NBD with arbitrary $k$ always has larger value of entropy as compared to the Poisson distribution with the same fixed value of $\langle n\rangle$. Their entropy becomes the same when $k\to \infty$, since NBD reduces to the Poisson distribution in that limit.

\section{Conclusions}

In this paper we have introduced and studied the Wehrl entropy of a hadron/nucleus defined through the QCD Husimi distribution. It quantifies the complexity of the multi-parton distribution in phase space $(b_\perp, k_\perp)$, and therefore it is a very interesting notion in the tomographic study of the nucleon. At small-$x$, our result parameterically agrees with the different definitions of (entanglement) entropy discussed in  \cite{Kutak:2011rb,Peschanski:2012cw,Kovner:2015hga}. Unlike in these previous works, however, the Wehrl entropy is given in terms of the gauge invariant matrix element of the quark and gluon field operators, and as such, it is not restricted to small-$x$ gluons.

The phenomenological implications of our result remain to be explored. It is often argued that the entropy is proportional to the final state multiplicity $dn/dY$, and the result $S_{W}\propto Q_s^2(Y) \sim e^{\alpha Y}$ appears to be consistent with the exponential growth of multiplicity with $Y\sim \ln s$. Now that we have a model-independent definition of entropy, such a correspondence can be pursued also at low-energy (large-$x$) and/or in quark-dominated processes.  Concerning the high-energy limit, our result in Section V suggests that the exponential growth will be tamed by the Pomeron loop effect, possibly leading to a nearly constant (in $Y$) multiplicity. But presumably this occurs at very high energy which has not been reached in modern accelerators yet. 



\section*{Acknowledgements}

Y.~Hatta thanks  Dmitri Kharzeev, Cedric Lorce and Hidekazu Tsukiji for helpful conversations. B.~Xiao and F.~Yuan  acknowledge interesting discussions with Volker Koch. Y.~Hatta and F.~Yuan thank the Central China Normal University for hospitality where this work has been completed. This material is based upon work supported by the U.S. Department of Energy, Office of Science, Office of Nuclear Physics, under contract number 
DE-AC02-05CH11231 and by the Natural Science Foundation of China (NSFC) under Grant No.~11575070. Y.~Hagiwara is supported by the JSPS KAKENHI Grant No. 17J08072.

\end{document}